\documentclass[aps,epsfig]{revtex4}
\usepackage{amssymb}
\usepackage{epsfig}
\usepackage{graphicx}
\usepackage{graphics}

\def\b{\begin{equation}}
\def\e{\end{equation}}
\def\balll{\begin{array}{lll}}
\def\ea{\end{array}}
\def\bea{\begin{eqnarray}}
\def\eea{\end{eqnarray}}

\newcommand{\be}{\begin{equation}}
\newcommand{\ee}{\end{equation}}
\newcommand{\bqn}{\begin{eqnarray}}
\newcommand{\eqn}{\end{eqnarray}}

\begin{document}

\title{Time evolution of a two-atom dressed entangled state in a cavity}
\author{E. R. Granhen$^{a,d}$, C. A. Linhares$^b$, A. P. C. Malbouisson$^a$
and J. M. C. Malbouisson$^c$}

\affiliation{$^a$Centro Brasileiro de Pesquisas F\'{\i}sicas/MCT,
22290-180, Rio de Janeiro, RJ, Brazil\\ $^b$Instituto de
F\'{\i}sica, Universidade do Estado do Rio de Janeiro, 20559-900,
Rio de Janeiro, RJ, Brazil\\ $^c$Instituto de F\'{\i}sica,
Universidade Federal da Bahia, 40210-340, Salvador, BA, Brazil\\
$^d$Faculdade de F\'{\i}sica, Universidade Federal do Par\'{a},
66075-110, Bel\'{e}m, PA, Brazil}

\date{\today }

\begin{abstract}
We study the time evolution of superposition of product states of
two dressed atoms in a spherical cavity in the extreme situations of
an arbitrarily large cavity (free space) and of a small one. In the
large-cavity case, the system dissipates, whereas, for the small
finite cavity, the system evolves in an oscillating way and never
completely decays. We also compute the von Neumann entropy for such
a system, a measurement of the degree of entanglement of the two
atoms, as the superposed state evolves in time. We find that this
entropy does not depend on time, nor on the size of the cavity.
\end{abstract}

\maketitle

\section{Introduction}

In the quantum mechanical description of multipartite systems, with
Hilbert spaces given by direct products of individual part spaces,
the superposition principle leads naturally to entangled states
which can not be written as single products of states of the
constituent parts; non interacting subsystems can thus share
entangled states that hold quantum correlations. Such quantum
entanglement carries nonlocal features which can be analyzed by
comparison with classical correlations~\cite{bel1,bel2}.

Entanglement is a quantum mechanical resource that plays a crucial
role in implementing teleportation of quantum states and in several
applications of quantum computation and quantum
information~\cite{ben1,bou1,NC}. Quantifying entanglement then
becomes an important issue which has been addressed in the
literature from a variety of
viewpoints~\cite{shi1,shi2,ben2,ved1,hil1,fer13,fer14}. For
bipartite systems, the measurement of entanglement is well
established, the von Neumann entropy of the reduced density matrix
providing the simplest measure of the degree of entanglement of a
given state. In this way, maximum entangled states has been
constructed for both boson~\cite{pli1,revze1} and
fermion~\cite{Esdras} bipartite systems.

In a recently reported experiment~\cite{Jost}, it is proven the
existence of deterministic entanglement of separated oscillators,
consisting of the vibrational states of two pairs of atomic ions in
different locations. They also demonstrate entanglement of the
internal states of an atomic ion with a distant mechanical
oscillator. The authors claim that such experiments may lead to the
generation of entangled states of mechanical oscillators in a larger
scale, in such a way as to provide tests for nonlocality in
mesoscopic systems. They also claim that these experiments could be
used to control quantum information processing based on trapped
atomic ions. Previously, an experiment was performed
in~\cite{knill}, using ultraviolet lasers to entangle two pairs of
beryllium ions in an electromagnetic trap. These authors also
cross-entangled the entangled pairs, that is, entangled each member
of the first pair with its correspondent in the second pair. Then
the first pair of ions was measured, and the results were used as an
indication of whether the unmeasured second pair was entangled.

In the present paper we study the time evolution of an entangled
two-atom state, in the presence of a force field. Our approach to
this problem makes use of the concept of dressed states. This
formalism, originally introduced in~\cite{adolfo1}, was already
employed to investigate several
situations~\cite{adolfo2,adolfo2a,adolfo3,adolfo4,adolfo5,linhares,termalizacao}.
It accounts for the fact that, for instance, a charged physical
particle is always coupled to the force (gauge) field; in other
words, it is always ``dressed'' by a cloud of quanta of the gauge
field. In general for a system of matter particles, the idea is that
the particles are coupled to an environment, which is usually
modeled in two equivalent ways: either to represent it by a free
field, as was done in Refs. \cite{zurek,paz}, or to consider the
environment as a reservoir composed of a large number of
noninteracting harmonic oscillators (see, for
instance,~\cite{ullersma,haake,caldeira,schramm}). In both cases,
exactly the same type of argument given above in the case of a
charged particle applies, with the appropriate changes, to such
systems. We may then speak of the ``dressing'' of the set of
particles by the ensemble of the harmonic modes of the environment.
It should be true in general for any system in which material
particles are coupled to an environment. In atomic physics, the
semiqualitative idea of a ``dressed atom'' has been largely employed
in studies involving the interaction of atoms and electromagnetic
fields~\cite {bouquinCohen}. In the realm of general physics, the
dressing of a matter particle by an environment has found an
application in describing the radiation damping of classical systems
\cite{petrosky}. Our dressed states can be viewed as a rigorous
version of these dressing procedures, in the context of the model
employed here.

We will consider our system in this paper as consisting of two
atoms, each one of them interacting independently inside a spherical
cavity with an environment provided by the harmonic modes of a
field. We take it as a bipartite system, each subsystem consisting
of one of the dressed atoms. We will consider a superposition of two
kinds of states: either all entities (both atoms and the field
modes) are in their ground states, or just one of the atoms lies in
its first excited state, the other one and all the field modes being
in their ground states. The analysis of the (reduced) density matrix
of the system leads to the computation of the von Neumann entropy,
which measures the degree of entanglement of the two atoms.

The dressing formalism for just one atom inside a cavity is briefly
reviewed in Section 2 in order to establish basic notation and
formulas for the time evolution of the states. In Section 3 the
formalism is generalized for the two-atom system and describe the
evolution of its density matrix, either in the case of a very large
cavity (with infinite radius, that is, free space) or of a small
cavity. The entanglement of the two atoms is discussed in Section 4.
Finally, in Section 5 we present our conclusions.

\section{A single dressed atom}

Before tackling the case of two atoms, it is convenient to reproduce
here the analysis of Ref.~\cite{linhares} for the simpler situation
of just one atom, dressed by its interaction with the environment
field. We present, in this section, a short review of the formalism
introduced in previous works.

We shall thus consider a atom in the harmonic approximation, coupled
linearly to an environment modeled by the infinite set of harmonic
modes of a scalar field, on the inside a spherical cavity. A
nonperturbative study of the time evolution of such a system is
implemented by means of \emph{dressed} states and
\emph{dressed} coordinates~\cite{adolfo1}. In particular, our
dressed states are \textit{not} the same as those currently employed
in the literature, usually associated to normal coordinates. Our
dressed states are given in terms of our dressed coordinates and
allow a rigorous study of the time evolution of quantum systems in
the context of the model employed here. The results we obtain by
these means are those expected on physical grounds, but contain
corrections with respect to the formulas obtained from perturbation
theory.

Let us start by considering an atom labeled $\lambda $, having
\emph{bare} frequency $\omega _\lambda $, linearly coupled to a
field described by $N$ ($ \rightarrow \infty $) other oscillators,
with frequencies $\omega _k$, $ k=1,2,\ldots ,N$. The whole system
is contained in a perfectly reflecting spherical cavity of radius
$R$, the free space corresponding to the limit $ R\rightarrow \infty
$. Hereafter, we shall refer to the harmonic oscillator as the
\textit{atom}, to distinguish it from the harmonic modes of the
environment. Denoting by $q_\lambda (t)$ ($p_\lambda (t)$) and
$q_k(t)$ ($ p_k(t)$) the coordinates (momenta) associated with the
atom and the field oscillators, respectively, the Hamiltonian of the
system is taken as
\begin{equation}
H_\lambda =\frac 12\left[ p_\lambda ^2+\omega _\lambda ^2q_\lambda
^2+\sum_{k=1}^N\left( p_k^2+\omega _k^2q_k^2\right) \right] -q_\lambda
\sum_{k=1}^N\eta _\lambda \omega _kq_k,  \label{Hamiltoniana}
\end{equation}
where $\eta _\lambda $ is a constant and the limit $N\rightarrow
\infty$ will be understood later on. The Hamiltonian
(\ref{Hamiltoniana}) can be turned to principal axis by means of a
point transformation,
\begin{equation}
q_{\mu(\lambda)} =\sum_{r_\lambda =0}^Nt_{\mu(\lambda)} ^{r_\lambda }Q_{r_\lambda
}\,\,,\,\,\,\,\,p_{\mu(\lambda)} = \sum_{r_\lambda =0}^Nt_{\mu(\lambda)} ^{r_\lambda
}P_{r_\lambda }\;,  \label{transf}
\end{equation}
where $\mu (\lambda)=(\lambda ,\{k\}),\,k=1,2,\ldots ,N$, and $r_\lambda =
0,\ldots ,N$ , performed by an orthonormal matrix $T=(t_{\mu(\lambda)}
^{r_\lambda })$. The subscripts $\mu =\lambda $ and $\mu =k$ refer
respectively to the atom and the harmonic modes of the field and
$r_\lambda $ refers to the normal modes. In terms of normal momenta
and coordinates, the transformed Hamiltonian reads
\begin{equation}
H_\lambda =\frac 12\sum_{r_\lambda =0}^N\left( P_{r_\lambda }^2+\Omega
_{r_\lambda }^2Q_{r_\lambda }^2\right) ,  \label{diagonal}
\end{equation}
where the $\Omega _{r_\lambda }$'s are the normal frequencies corresponding
to the collective \textit{stable} oscillation modes of the coupled system.

Using the coordinate transformation $q_{\mu(\lambda)} =\sum_{r_\lambda }t_{\mu(\lambda)}
^{r_\lambda }Q_{r_\lambda }$ in the equations of motion and explicitly
making use of the normalization condition
\begin{equation}
\sum_{\mu =0}^N\left( t_{\mu(\lambda)} ^{r_\lambda }\right) ^2=1,  \label{normcond}
\end{equation}
we get
\begin{equation}
t_k^{r_\lambda }=\frac{\eta _\lambda \omega _k}{\omega _k^2-\Omega
_{r_\lambda }^2}t_\lambda ^{r_\lambda }\;,\;\;t_\lambda ^{r_\lambda
} = \left[ 1+\sum_{k=1}^N\frac{\eta _\lambda ^2\omega _k^2}{(\omega
_k^2-\Omega _{r_\lambda }^2)^2}\right] ^{-\frac 12},  \label{tkrg1}
\end{equation}
with the condition
\begin{equation}
\omega _\lambda ^2-\Omega _{r_\lambda }^2=\sum_{k=1}^N\frac{\eta _\lambda
^2\omega _k^2}{\omega _k^2-\Omega _{r_\lambda }^2}.  \label{Nelson1}
\end{equation}
The right-hand side of equation~(\ref{Nelson1}) diverges in the
limit $ N\rightarrow \infty $. Defining the counterterm $\delta
\omega ^2=N\eta _\lambda ^2$, it can be rewritten in the form
\begin{equation}
\omega _\lambda ^2-\delta \omega ^2-\Omega _{r_\lambda }^2=\eta _\lambda
^2\Omega _{r_\lambda }^2\sum_{k=1}^N\frac 1{\omega _k^2-\Omega _{r_\lambda
}^2}.  \label{Nelson2}
\end{equation}

Equation (\ref{Nelson2}) has $N+1$ solutions, corresponding to the
$N+1$ normal collective modes. It can be shown~\cite{adolfo1} that
if $\omega _\lambda ^2>\delta \omega ^2$, all possible solutions for
$\Omega ^2$ are positive, physically meaning that the system
oscillates harmonically in all its modes. On the other hand, when
$\omega _\lambda ^2<\delta \omega ^2$, one of the solutions is
negative and so no stationary configuration is allowed.

Therefore, we just consider the situation in which all normal modes are
harmonic, which corresponds to the first case above, $\omega _\lambda
^2>\delta \omega ^2$, and define the \textit{renormalized} frequency
\begin{equation}
{\bar{\omega}}_\lambda ^2=\lim_{N\rightarrow \infty }\left( \omega _\lambda
^2-N\eta _\lambda ^2\right) ,  \label{omegabarra}
\end{equation}
following the pioneering work of Ref.~\cite{Thirring}. In the limit
$ N\rightarrow \infty $, equation (\ref{Nelson2}) becomes
\begin{equation}
{\bar{\omega}}_\lambda ^2-\Omega ^2=\eta _\lambda
^2\sum_{k=1}^\infty \frac{ \Omega ^2}{\omega _k^2-\Omega ^2}.
\label{Nelson3}
\end{equation}
We see that, in this limit, the above procedure is exactly the
analogous of mass renormalization in quantum field theory: the
addition of a counterterm $ -N\eta _\lambda ^2q_\lambda ^2$
($N\rightarrow \infty $) allows one to compensate the infinity of
$\omega _\lambda ^2$ in such a way as to leave a finite, physically
meaningful, renormalized frequency ${\bar{\omega}} _\lambda $.

To proceed, we take the constant $\eta _\lambda $ as
\begin{equation}
\eta _\lambda =\sqrt{\frac{4g_\lambda \Delta \omega }\pi },  \label{eta}
\end{equation}
where $\Delta \omega $ is the interval between two neighboring field
frequencies and $g$ is the coupling constant with dimension of frequency.
The environment frequencies $\omega _k$ can be written in the form
\begin{equation}
\omega _k=k\frac{\pi c}R,\;\;\;\;k=1,2,\ldots ,  \label{discreto}
\end{equation}
and, so, $\Delta \omega =\pi c/R$. Then, using the identity
\begin{equation}
\sum_{k=1}^\infty \frac 1{k^2-u^2}=\frac 12\left[ \frac 1{u^2}-\frac \pi
u\cot \left( \pi u\right) \right] ,  \label{id4}
\end{equation}
equation~(\ref{Nelson3}) can be written in closed form:
\begin{equation}
\cot \left( \frac{R\Omega }c\right) =\frac \Omega {2g_\lambda
}+\frac c{R\Omega }\left( 1-\frac{R{\bar{\omega}}_\lambda
^2}{2g_\lambda c}\right) . \label{eigenfrequencies1}
\end{equation}
The elements of the transformation matrix, turning the atom--field
system to principal axis, are obtained in terms of the physically
meaningful quantities $\Omega _{r_\lambda }$ and
${\bar{\omega}}_\lambda $ after some rather long but straightforward
manipulations \cite{adolfo1}. They read
\begin{eqnarray}
t_\lambda ^{r_\lambda } &=&\frac{\eta _\lambda \Omega _{r_\lambda
}}{\sqrt{ \left( \Omega _{r_\lambda }^2-{\bar{\omega}}_\lambda
^2\right) ^2+\frac{\eta _\lambda ^2}2\left( 3\Omega _{r_\lambda
}^2-{\bar{\omega}}_\lambda ^2\right)
+4g_\lambda ^2\Omega _{r_\lambda }^2}},  \label{t0r2} \\
t_k^{r_\lambda } &=&\frac{\eta _\lambda \omega _k}{\omega _k^2-\Omega
_{r_\lambda }^2}t_\lambda ^{r_\lambda }.  \label{t0r21}
\end{eqnarray}

Let us now consider the eigenstates of the system atom($\lambda$)-field, $\left|
l_{\lambda},l_1,l_2,...\right\rangle $, represented by the normalized
eigenfunctions, written in terms of the normal coordinates $\{Q_{r_\lambda
}\}$,
\begin{equation}
\phi _{l_\lambda l_1l_2...}(Q,t)=\prod_s\left[
\sqrt{\frac{2^{l_{s_{\lambda}}}}{l_{s_{\lambda}}!}} H_{l_{s_{\lambda}}}\left( \sqrt{\frac{\Omega
_{s_{\lambda}}}\hbar }Q_{s_{\lambda}}\right) \right] \Gamma _0^\lambda \,e^{-i\sum_{s_{\lambda}}\left(
l_{s_{\lambda}}+\frac 12\right) \Omega _{s_{\lambda}}t}, \label{autofuncoes}
\end{equation}
where $H_{l_{s_{\lambda}}}$ stands for the $l_{s_{\lambda}}$-th Hermite polynomial and
\begin{equation}
\Gamma _0^\lambda = {\mathcal{N}}_{\lambda}e^{-\sum_s\frac{\Omega_{s_{\lambda}}Q_{s_{\lambda}}^2}{2}}
\label{gamazero}
\end{equation}
is the normalized vacuum eigenfunction, ${\mathcal{N}}_{\lambda}$ being the normalization factor. 

We introduce \textit{dressed} coordinates $q_\lambda ^{\prime
}$ and $ \{q_i^{\prime }\}$ for the \textit{dressed }atom and the
\textit{dressed} field, respectively, defined by
\begin{equation}
\sqrt{\bar{\omega}_{\mu(\lambda)} }q_{\mu(\lambda)} ^{\prime }=\sum_{r_\lambda }t_{\mu(\lambda)}
^{r_\lambda } \sqrt{\Omega _{r_\lambda }}Q_{r_\lambda },
\label{qvestidas1}
\end{equation}
where ${\bar{\omega}}_{\mu(\lambda)} =\{{\bar{\omega}}_\lambda ,\;\omega _i\}$.
In terms of the dressed coordinates, we define for a fixed
instant, $t=0$, \textit{dressed} states, $\left| \kappa _\lambda
,\kappa _1,\kappa _2,\cdots \right\rangle $ by means of the complete
orthonormal set of functions~\cite{adolfo1}
\begin{equation}
\psi _{\kappa _\lambda \kappa _1...}(q^{\prime })=\prod_{\mu(\lambda)} \left[
\sqrt{ \frac{2^{\kappa _{\mu(\lambda)} }}{\kappa _{\mu(\lambda)} !}}H_{\kappa _{\mu(\lambda)} }\left(
\sqrt{\frac{ \bar{\omega}_{\mu(\lambda)} }\hbar }q_{\mu(\lambda)} ^{\prime }\right)
\right] \Gamma _0^\lambda , \label{ortovestidas1}
\end{equation}
where $\mu(\lambda)$ labels collectively the dressed atom $\lambda $ and the
field modes, $1,2,3,\ldots $, $q_{\mu(\lambda)} ^{\prime } = q_\lambda
^{\prime },\,\left\{ q_i^{\prime }\right\} $. The ground state
$\Gamma _0^\lambda $ in the above equation is the same as in
equation~(\ref{autofuncoes}). The invariance of the ground state is
due to our definition of dressed coordinates given by
equation~(\ref{qvestidas1}). Notice that the introduction of the dressed coordinates
implies, differently from the bare vacuum, the stability of the
dressed vacuum state since, by construction, it is identical to the
ground state of the interacting Hamiltonian (\ref{diagonal}).
Each function $\psi _{\kappa _\lambda
\kappa _1...}(q^{\prime })$ describes a state in which the dressed
oscillator $q_\mu ^{\prime }$ is in its $\kappa _{\mu(\lambda)} $-th excited
state. 

Let us consider the particular dressed state $\left| \Gamma _1^{\mu
(\lambda )}(0)\right\rangle $ at $t=0$, represented by the wave
function $\psi _{00\cdots 1(\mu )0\cdots }(q^{\prime })$. It
describes the configuration in which \textit{only} the $\mu $-th
dressed oscillator is in the \textit{first} excited level, all other
 being in their ground states. As shown in
Ref.~\cite{adolfo1}, the time evolution of the state $\left| \Gamma
_1^{\mu (\lambda )}\right\rangle $ is given by
\begin{equation}
\left| \Gamma _1^{\mu (\lambda )}(t)\right\rangle =\sum_\nu f_{\mu \nu
}(t)\left| \Gamma _1^{\nu (\lambda )}(0)\right\rangle ,
\label{ortovestidas51}
\end{equation}
where $\mu (\lambda ),\nu (\lambda )=\lambda ,\{i\}$, with $\{i\}$
referring to the field modes, and
\begin{equation}
f_{\mu \nu }(t)=\sum_st_{\mu} ^st_\nu ^se^{-i\Omega _st}.
\label{ortovestidas5}
\end{equation}
Moreover, it can be shown that, for all $\mu $,
\begin{equation}
\sum_\nu \left| f_{\mu \nu }(t)\right| ^2=1,  \label{probabilidade}
\end{equation}
which allows to interpret the coefficients $f_{\mu \nu }(t)$ as
probability amplitudes; for example, $f_{\lambda \lambda }(t)$ is
the probability amplitude that, if the dressed atom is in the first
excited state at $t=0$, it remains excited at time $t$, while
$f_{\lambda i}(t)$ represents the probability amplitude that the
$i$-th dressed harmonic mode of the field be at the first excited
level.

\section{Time evolution of a dressed two-atom state}

We now consider a bipartite system composed of two subsystems,
$\mathcal{A}$ and $\mathcal{B}$; the subsystems consist respectively
of dressed atoms $A$ and $B$, in the sense defined in the preceding
section, the whole system being contained in a perfectly reflecting
sphere of radius $R$. Let us consider the eigenstates of the
subsystems $\mathcal{A}$ and $\mathcal{B}$ with $\lambda =A,B$
labeling the quantities referring to the subsystems.

We consider the Hilbert space spanned by the dressed Fock-like states,
\begin{equation}
\left| \Gamma _{n_Ak_1k_2\cdots ;\,n_Bq_1q_2\cdots }^{(AB)}\right\rangle
\equiv |n_A,k_1,k_2,\ldots ;\,n_B,q_1,q_2,\ldots \rangle =\left| \Gamma
_{n_A,k_1,k_2,\ldots }^A\right\rangle \otimes \left| \Gamma
_{n_B,q_1,q_2,\ldots }^B\right\rangle ,
\end{equation}
in which the dressed atom $A$ is at the $n_A$ excited level and the
atom $B$ is at the $n_B$ excited level; the (doubled) dressed modes
of the field are at the $k_1,k_2,\ldots $, $q_1,q_2,\ldots $ excited
levels. Using this definition, let us consider at time $t=0$, a
family of entangled states of the bipartite system given by
\begin{eqnarray}
\left| \Psi \right\rangle & = &\sqrt{\xi }\,\left| \Gamma
_{1(A)00\cdots ;0(B)00\cdots }^{(AB)}(0)\right\rangle +\sqrt{1-\xi
}\,e^{i\phi }\,\left| \Gamma _{0(A)00\cdots ;\,1(B)00\cdots
}^{(AB)}(0)\right\rangle \nonumber \\
&=&\sqrt{\xi }\,\left| 1_A,0,0,\cdots ;0_B,0,0,\cdots \right\rangle
+\sqrt{ 1-\xi }\,e^{i\phi }\,\left| 0_A,0,0,\cdots ;\,1_B,0,0,\cdots
\right\rangle , \label{definition-entangled-state}
\end{eqnarray}
where $0<\xi <1$. In equation (\ref{definition-entangled-state}), $\left|
\Gamma _{1(A)0(B)00\cdots }^{(AB)}(0)\right\rangle $ and $\left| \Gamma
_{0(A)1(B)00\cdots }^{(AB)}(0)\right\rangle $ stand respectively for the
states in which the dressed atom $A$ ($B$) is at the first level, the
dressed atom $B$ ($A$) and all the field modes being in the ground state.
They are
\begin{equation}
\left| \Gamma _{1(A)0(B)00\cdots }^{(AB)}(0)\right\rangle =\left| \Gamma
_{100\cdots }^A(0)\right\rangle \otimes \left| \Gamma _{000\cdots
}^B(0)\right\rangle
\end{equation}
and
\begin{equation}
\left| \Gamma _{0(A)1(B)00\cdots }^{(AB)}(0)\right\rangle =\left| \Gamma
_{000\cdots }^A\right\rangle \otimes \left| \Gamma _{100\cdots
}^B(0)\right\rangle .
\end{equation}

The density matrix at $t=0$ is
\begin{eqnarray}
\varrho (0) &=&\left| \Psi \right\rangle \left\langle \Psi \right|
\nonumber
\\
&=&\xi \left| 1_A,0,0,\cdots ;0_B,0,0,\cdots \right\rangle \left\langle
1_A,0,0,\cdots ;0_B,0,0,\cdots \right|  \nonumber \\
&&+\left( 1-\xi \right) \left| 0_A,0,0,\cdots ;\,1_B,0,0,\cdots
\right\rangle \left\langle 0_A,0,0,\cdots ;\,1_B,0,0,\cdots \right|
\nonumber \\
&&+\sqrt{\xi (1-\xi )}e^{-i\phi }\left| 1_A,0,0,\cdots ;0_B,0,0,\cdots
\right\rangle \left\langle 0_A,0,0,\cdots ;\,1_B,0,0,\cdots \right|
\nonumber \\
&&+\sqrt{\xi (1-\xi )}e^{i\phi }\left| 0_A,0,0,\cdots ;\,1_B,0,0,\cdots
\right\rangle \left\langle 1_A,0,0,\cdots ;0_B,0,0,\cdots \right| ,
\end{eqnarray}

At time $t$, the state of the system is described by the density matrix
\begin{equation}
\varrho (t)=e^{-iHt}\left| \Psi \right\rangle \left\langle \Psi \right|
e^{iHt},  \label{density-matrix}
\end{equation}
where $H$ is the Hamiltonian of the whole system, such that
\[
e^{-iHt}=e^{-iH_At}\otimes e^{-iH_Bt}
\]
and $H_A$ and $H_B$ are the Hamiltonian $H_\lambda $ of equations (\ref
{Hamiltoniana}) or (\ref{diagonal}). We then obtain
\begin{eqnarray}
\varrho (t) &=&\xi \left( \left| \Gamma _{100\cdots }^A(t)\right\rangle
\left\langle \Gamma _{100\cdots }^A(t)\right| \right) \otimes \left( \left|
\Gamma _{000\cdots }^B\right\rangle \left\langle \Gamma _{000\cdots
}^B\right| \right)  \nonumber \\
&&+\left( 1-\xi \right) \left( \left| \Gamma _{000\cdots }^A\right\rangle
\left\langle \Gamma _{000\cdots }^A\right| \right) \otimes \left( \left|
\Gamma _{100\cdots }^B(t)\right\rangle \left\langle \Gamma _{100\cdots
}^B(t)\right| \right)  \nonumber \\
&&+\sqrt{\xi (1-\xi )}e^{i\phi }\left( \left| \Gamma _{000\cdots
}^A\right\rangle \left\langle \Gamma _{100\cdots }^A(t)\right| \right)
\otimes \left( \left| \Gamma _{100\cdots }^B(t)\right\rangle \left\langle
\Gamma _{000\cdots }^B\right| \right)  \nonumber \\
&&+\sqrt{\xi (1-\xi )}e^{-i\phi }\left( \left| \Gamma _{100\cdots
}^A(t)\right\rangle \left\langle \Gamma _{000\cdots }^A\right| \right)
\otimes \left( \left| \Gamma _{000\cdots }^B\right\rangle \left\langle
\Gamma _{100\cdots }^B(t)\right| \right) ,
\end{eqnarray}
where the states $\left| \Gamma _{000\cdots }^A\right\rangle $, $\left|
\Gamma _{000\cdots }^B\right\rangle $ are stationary and the states $\left|
\Gamma _{100\cdots }^A(t)\right\rangle $, $\left| \Gamma _{100\cdots
}^B(t)\right\rangle \,$evolve according to equation (\ref{ortovestidas51}).

In order to investigate how the superposed states evolve in time, we
shall consider the reduced density matrix obtained by tracing over
all the degrees of freedom associated with the field. The
computation is analogous to the one presented in Ref.
\cite{linhares}. After taking the trace, the density matrix has the
indices referring to the 2-atom states. Explicitly, we have
\begin{eqnarray}
\rho _{n_{\mathcal{A}}n_{\mathcal{B}}}^{m_{\mathcal{A}}m_{\mathcal{B}}}(t)
&=&\xi \sum_{\{k_i=1\}}^\infty \left\langle n_{\mathcal{A}},k_1,k_2,\ldots
|\Gamma _{100\cdots }^A(t)\right\rangle \left\langle \Gamma _{100\cdots
}^A(t)|m_{\mathcal{A}},k_1,k_2,\ldots \right\rangle   \nonumber \\
&&\qquad \times \sum_{\{q_i=1\}}^\infty \left\langle n_{\mathcal{B}
},q_1,q_2,\ldots |\Gamma _{000\cdots }^B\right\rangle \left\langle
\Gamma
_{000\cdots }^B|m_{\mathcal{B}},q_1,q_2,\ldots \right\rangle   \nonumber \\
&&+(1-\xi )\sum_{\{k_i=1\}}^\infty \left\langle n_{\mathcal{A}
},k_1,k_2,\ldots |\Gamma _{000\cdots }^A\right\rangle \left\langle
\Gamma
_{000\cdots }^A|m_{\mathcal{A}},k_1,k_2,\ldots \right\rangle   \nonumber \\
&&\qquad \times \sum_{\{q_i=1\}}^\infty \left\langle n_{\mathcal{B}
},q_1,q_2,\ldots |\Gamma _{100\cdots }^B(t)\right\rangle
\left\langle \Gamma _{100\cdots
}^B(t)|m_{\mathcal{B}},q_1,q_2,\ldots \right\rangle   \nonumber
\\
&&+\sqrt{\xi (1-\xi )}e^{i\phi }\sum_{\{k_i=1\}}^\infty \left\langle
n_{ \mathcal{A}},k_1,k_2,\ldots |\Gamma _{000\cdots }^A\right\rangle
\left\langle \Gamma _{100\cdots
}^A(t)|m_{\mathcal{A}},k_1,k_2,\ldots
\right\rangle   \nonumber \\
&&\qquad \times \sum_{\{q_i=1\}}^\infty \left\langle n_{\mathcal{B}
},q_1,q_2,\ldots |\Gamma _{100\cdots }^B(t)\right\rangle
\left\langle \Gamma
_{000\cdots }^B|m_{\mathcal{B}},q_1,q_2,\ldots \right\rangle   \nonumber \\
&&+\sqrt{\xi (1-\xi )}e^{-i\phi }\sum_{\{k_i=1\}}^\infty
\left\langle n_{ \mathcal{A}},k_1,k_2,\ldots |\Gamma _{100\cdots
}^A(t)\right\rangle \left\langle \Gamma _{000\cdots
}^A|m_{\mathcal{A}},k_1,k_2,\ldots
\right\rangle   \nonumber \\
&&\qquad \times \sum_{\{q_i=1\}}^\infty \left\langle n_{\mathcal{B}
},q_1,q_2,\ldots |\Gamma _{000\cdots }^B\right\rangle \left\langle
\Gamma _{100\cdots }^B(t)|m_{\mathcal{B}},q_1,q_2,\ldots
\right\rangle .
\end{eqnarray}
In the above expression we have typically
\begin{equation}
\left\langle n_{\mathcal{A}},k_1,k_2,\ldots |\Gamma _{000\cdots
}^A\right\rangle =\delta _{k_10}\delta _{k_20}\cdots
\end{equation}
and
\begin{eqnarray}
\left\langle n_{\mathcal{A}},k_1,k_2,\ldots |\Gamma _{100\cdots
}^A(t)\right\rangle  &=&\sum_\nu f_{A\nu }(t)\left\langle
n_{\mathcal{A}
},k_1,k_2,\ldots |\Gamma _{100\cdots }^\nu (0)\right\rangle   \nonumber \\
&=&f_{AA}(t)\left\langle n_{\mathcal{A}},k_1,k_2,\ldots |\Gamma
_{100\cdots }^A(0)\right\rangle +\sum_{i=1}^\infty
f_{Ai}(t)\left\langle n_{\mathcal{A}
},k_1,k_2,\ldots |\Gamma _{100\cdots }^i(0)\right\rangle   \nonumber \\
&=&f_{AA}(t)\delta _{n_{\mathcal{A}}1}\delta _{k_10}\delta _{k_20}\cdots
+\sum_{i=1}^\infty f_{Ai}(t)\delta _{n_{\mathcal{A}}0}\delta _{k_10}\cdots
\delta _{k_i1}\cdots
\end{eqnarray}
so that the sums in the elements of the reduced density matrix are
of one of the types below:
\begin{eqnarray}
\sum_{k_1,k_2,\ldots }\left\langle n_{\mathcal{A}},k_1,k_2,\ldots
|\Gamma _{000\cdots }^A\right\rangle \left\langle \Gamma _{000\cdots
}^A|m_{\mathcal{ A}},k_1,k_2,\ldots \right\rangle  &=&\delta
_{n_{\mathcal{A}}0}\delta _{m_{ \mathcal{A}}0}\sum_{k_1}\delta
_{k_10}\delta _{k_10}\sum_{k_2}\delta
_{k_20}\delta _{k_20}\cdots   \nonumber \\
&=&\delta _{n_{\mathcal{A}}0}\delta _{m_{\mathcal{A}}0},
\end{eqnarray}
\begin{eqnarray}
&&\sum_{k_1,k_2,\ldots }\left\langle n_{\mathcal{A}},k_1,k_2,\ldots
|\Gamma _{000\cdots }^A\right\rangle \left\langle \Gamma _{100\cdots
}^A(t)|m_{
\mathcal{A}},k_1,k_2,\ldots \right\rangle   \nonumber \\
&=&\delta _{n_{\mathcal{A}}0}\left[ f_{AA}^{*}(t)\delta
_{m_{\mathcal{A} }1}\sum_{k_1}\delta _{k_10}\delta
_{k_10}\sum_{k_2}\delta _{k_20}\delta _{k_20}\cdots
+\sum_if_{Ai}^{*}(t)\sum_{k_1}\delta _{k_10}\delta _{k_10}\cdots
\sum_{k_i}\delta _{k_i0}\delta _{k_i1}\cdots \right]
\nonumber \\
&=&f_{AA}^{*}(t)\delta _{n_{\mathcal{A}}0}\delta _{m_{\mathcal{A}}1},
\end{eqnarray}
and
\begin{eqnarray}
&&\sum_{k_1,k_2,\ldots }\left\langle n_{\mathcal{A}},k_1,k_2,\ldots
|\Gamma _{100\cdots }^A(t)\right\rangle \left\langle \Gamma
_{100\cdots }^A(t)|m_{
\mathcal{A}},k_1,k_2,\ldots \right\rangle   \nonumber \\
&=&\sum_{k_1,k_2,\ldots }\left[ f_{AA}(t)\delta _{n_{\mathcal{A}}1}\delta
_{k_10}\delta _{k_20}\cdots +\sum_if_{Ai}(t)\delta _{n_{\mathcal{A}}0}\delta
_{k_10}\cdots \delta _{k_i1}\cdots \right]   \nonumber \\
&&\qquad \times \left[ f_{AA}^{*}(t)\delta _{m_{\mathcal{A}}1}\delta
_{k_10}\delta _{k_20}\cdots +\sum_jf_{Aj}^{*}(t)\delta
_{m_{\mathcal{A}
}0}\delta _{k_10}\cdots \delta _{k_i1}\cdots \right]   \nonumber \\
&=&\left| f_{AA}\right| ^2\delta _{n_{\mathcal{A}}1}\delta
_{m_{\mathcal{A} }1}+\sum_i\left| f_{Ai}\right| ^2\delta
_{n_{\mathcal{A}}0}\delta _{m_{ \mathcal{A}}0}.
\end{eqnarray}
Collecting all these expressions and their analogues for the atom $B$ into
the elements of the reduced density matrix we finally obtain
\begin{eqnarray}
\rho
_{n_{\mathcal{A}}n_{\mathcal{B}}}^{m_{\mathcal{A}}m_{\mathcal{B}}}(t)
&=&\xi \left[ \left| f_{AA}(t)\right| ^2\delta
_{n_{\mathcal{A}}1}\delta _{m_{\mathcal{A}}1}+\sum_i\left|
f_{Ai}(t)\right| ^2\delta _{n_{\mathcal{A} }0}\delta
_{m_{\mathcal{A}}0}\right] \delta _{n_{\mathcal{B}}0}\delta _{m_{
\mathcal{B}}0}  \nonumber \\
&&+(1-\xi )\delta _{n_{\mathcal{A}}0}\delta _{m_{\mathcal{A}}0}\left[ \left|
f_{BB}(t)\right| ^2\delta _{n_{\mathcal{B}}1}\delta _{m_{\mathcal{B}%
}1}+\sum_i\left| f_{Bi}(t)\right| ^2\delta _{n_{\mathcal{B}}0}\delta
_{m_{
\mathcal{B}}0}\right]   \nonumber \\
&&+\sqrt{\xi (1-\xi )}e^{i\phi }f_{AA}^{*}(t)f_{BB}(t)\delta
_{n_{\mathcal{A} }0}\delta _{m_{\mathcal{A}}1}\delta
_{n_{\mathcal{B}}1}\delta _{m_{\mathcal{B
}}0}  \nonumber \\
&&+\sqrt{\xi (1-\xi )}e^{-i\phi }f_{AA}(t)f_{BB}^{*}(t)\delta
_{n_{\mathcal{A }}1}\delta _{m_{\mathcal{A}}0}\delta
_{n_{\mathcal{B}}0}\delta _{m_{\mathcal{ B}}1}.
\end{eqnarray}
That is, the nonvanishing elements are given by
\begin{eqnarray}
\rho _{0_{\mathcal{A}}0_{\mathcal{B}}}^{0_{\mathcal{A}}0_{\mathcal{B}}}(t)
&=&1-\xi \left| f_{AA}(t)\right| ^2-(1-\xi )\left| f_{BB}(t)\right| ^2,
\nonumber \\
\rho _{0_{\mathcal{A}}1_{\mathcal{B}}}^{0_{\mathcal{A}}1_{\mathcal{B}}}(t)
&=&(1-\xi )\left| f_{BB}(t)\right| ^2,  \nonumber \\
\rho
_{1_{\mathcal{A}}0_{\mathcal{B}}}^{1_{\mathcal{A}}0_{\mathcal{B}}}(t)
&=&\xi \left| f_{AA}(t)\right| ^2, \label{elementos}  \\
\rho_{0_{\mathcal{A}}1_{\mathcal{B}}}^{1_{\mathcal{A}}0_{\mathcal{B}}}(t)
&=&\sqrt{\xi (1-\xi )}e^{i\phi }f_{AA}^{*}(t)f_{BB}(t),  \nonumber \\
\rho _{1_{\mathcal{A}}0_{\mathcal{B}}}^{0_{\mathcal{A}}1_{\mathcal{B}}}(t)
&=&\sqrt{\xi (1-\xi )}e^{-i\phi }f_{AA}(t)f_{BB}^{*}(t),  \nonumber
\end{eqnarray}
where equation (\ref{probabilidade}) was used. We check immediately that the
trace of this reduced density matrix is one,
\begin{equation}
\rho
_{0_{\mathcal{A}}0_{\mathcal{B}}}^{0_{\mathcal{A}}0_{\mathcal{B}}}+\rho
_{0_{\mathcal{A}}1_{\mathcal{B}}}^{0_{\mathcal{A}}1_{\mathcal{B}}}+\rho
_{1_{
\mathcal{A}}0_{\mathcal{B}}}^{1_{\mathcal{A}}0_{\mathcal{B}}}(t)+\rho
_{1_{
\mathcal{A}}1_{\mathcal{B}}}^{1_{\mathcal{A}}1_{\mathcal{B}}}(t)=1.
\label{traco11}
\end{equation}
This property ensures that $\rho $ represents physical states of the system.
Also, we see that Tr$\left[ \rho ^2\right] \neq 1$ and therefore, the
superposed states are not pure. The degree of impurity of a quantum state
can be quantified by the departure from the idempotency property. In the
present case:
\begin{eqnarray}
D(t,\xi ) &=&1-\text{Tr}\left[ \rho ^2\right]   \nonumber \\
&=&2\left( \xi \left| f_{AA}(t)\right| ^2+(1-\xi )\left|
f_{BB}(t)\right| ^2\right)  - 2\left( \xi \left| f_{AA}(t)\right|
^2+(1-\xi )\left| f_{BB}(t)\right| ^2\right) ^2.  \label{Dezao}
\end{eqnarray}

In the remainder of this section we consider the two atoms as identical and,
accordingly, we adopt the subscript $0$ for both of them, $\lambda
=A=B\equiv 0$; we also define
\begin{equation}
g_A=g_B\equiv g\,;\;\;\eta _A=\eta _B\equiv \eta
\,;\;\;{\bar{\omega}}_A={ \bar{\omega}}_B\equiv
\bar{\omega}\,;\;\;f_{AA}(t)=f_{BB}(t)\equiv f_{00}(t).
\end{equation}
In this case, the matrix elements in equations~(\ref{elementos})
simplify and, from equation~(\ref{Dezao}), we see that the degree of
impurity becomes independent of the superposition parameter $\xi $:
\begin{equation}
D(t,\xi )=2\left| f_{00}(t)\right| ^2(1-\left| f_{00}(t)\right| ^2).
\label{Dezao1}
\end{equation}
In order to pursue the study of the time evolution of the
superposition of the two-atom states, we have to determine the
behavior of $f_{00}(t)$. We shall analyze it in the situations of a
very large cavity (free space) and of a small one.

\subsection{The limit of an arbitrarily large cavity}

We start from the matrix element $t_\lambda ^{r_\lambda }$ in
equation~(\ref{t0r2}) and consider an arbitrarily large radius $R$
for the cavity. The two atoms behave independently from each other,
so let us focus on just one of them, either the atom $A$ or the atom
$B$. Remembering that $\eta =\sqrt{4gc/R}$, we have
\begin{equation}
\lim_{R\rightarrow \infty }t_0^r=\lim_{R\rightarrow \infty
}\frac{\sqrt{ 4g/\pi }\Omega \sqrt{\pi c/R}}{\sqrt{\left( \Omega
^2-\bar{\omega}^2\right) ^2+4g^2\Omega ^2}}.  \label{limite-t-r-0}
\end{equation}
In this limit, $\Delta \omega = \pi c/R\rightarrow d\omega = d\Omega
$ and the sum in the definition of $f_{00}(t)$, equation
(\ref{ortovestidas5}), becomes an integral, so that
\begin{equation}
f_{00}(t)=\frac{4g}\pi \int_0^\infty d\Omega \frac{\Omega
^2e^{-i\Omega t}}{ \left( \Omega ^2-\bar{\omega}^2\right)
^2+4g^2\Omega ^2}. \label{integral-f-00}
\end{equation}

We then proceed as in~\cite{linhares}. We define a parameter $\kappa
=\sqrt{ \bar{\omega}^2-g^2}$ and consider whether $\kappa ^2\geq 0$
or $\kappa ^2<0$ , for which $\kappa ^2\gg 0$ and $\kappa ^2\ll 0$
correspond respectively to weak ($g\ll \bar{\omega}_A$) and strong
($g\gg \bar{\omega}_A$) coupling of the atoms with the environmment.
For definiteness we consider in the following the weak-coupling
regime. We get in this case \cite{linhares}
\begin{equation}
f_{00}(t)=e^{-gt}\left[ \cos \kappa t-\frac g\kappa \sin \kappa t\right]
+iG\left( t;\bar{\omega},g\right) ,  \label{eq27}
\end{equation}
where the function $G(t;\bar{\omega},g)$ is given by
\begin{equation}
G(t;\bar{\omega},g)=-\frac{4g}\pi \int_0^\infty dx\frac{x^2\sin xt}{\left(
x^2-\bar{\omega}^2\right) ^2+4g^2x^2}.  \label{J}
\end{equation}
For large times, the quantity $\left| f_{00}(t)\right| ^2$ is given
by~\cite{linhares}
\begin{equation}
\left| f_{00}(t)\right| ^2\approx e^{-2gt}\left[ \cos
\bar{\omega}t-\frac g{ \bar{\omega}}\sin \bar{\omega}t\right]
^2+\frac{64g^2}{\bar{\omega}^8t^6}. \label{FAA2}
\end{equation}
As $t\rightarrow \infty $, we see that the expression for $\left|
f_{00}(t)\right| ^2$ go to zero.

\subsection{Small cavity}

For a finite (small) cavity, the spectrum of eigenfrequencies is
discrete, $ \Delta \omega $ is large, and so the approximation made
in the case of large cavity does not apply; no analytical result can
be obtained for $f_{00}(t)$ in this case. For a sufficiently small
cavity, the frequencies $\Omega _{r} $ can be determined by
following the steps described in~\cite{linhares}. Let us label the
eigenfrequencies as $\Omega _0$, $\left\{ \Omega _k\right\} $, $
k=1,2,\ldots $ Then, defining the dimensionless parameter
\begin{equation}
\delta =\frac g{\Delta \omega }=\frac{gR}{\pi c},  \label{delta}
\end{equation}
we rewrite equation~(\ref{eigenfrequencies1}) in the form
\begin{equation}
\cot \left( \frac{R\Omega _{r}}c\right) =\frac{\Omega _{r}
}{2g}+\frac c{R\Omega _{r}}\left( 1-\frac{R\bar{\omega}^2}{2gc}
\right) .  \label{cota}
\end{equation}
Taking $\delta \ll 1$, which corresponds to $R\ll \pi c/g$ (a small
cavity), it is shown in \cite{linhares} that, for $k=1,2,\ldots $,
the solutions are
\begin{equation}
\Omega _k\approx \frac g\delta \left( k+\frac{2\delta }{\pi k}\right) .
\label{OmegaK}
\end{equation}
If we further impose that $\delta <2g^2/\pi \bar{\omega}^2$, a condition
compatible with $\delta \ll 1$, then $\Omega _0$ is found to be very close
to $\bar{\omega}$, that is,
\begin{equation}
\Omega _0\approx \bar{\omega}\left( 1-\frac{\pi \delta }3\right) .
\label{Omega0}
\end{equation}

To determine $f_{00}(t)$, we have to calculate the square of the matrix
elements $\left( t_0^0\right) ^2$ and $\left( t_k^0\right) ^2$. They are
given, to first order in $\delta $, by
\begin{equation}
\left( t_0^0\right) ^2\approx \left( 1+\frac{2\pi \delta }3\right)
^{-1};\qquad \left( t_k^0\right) ^2\approx \frac 4{k^2}\frac \delta \pi
(t_0^0)^2.  \label{elem-t-k-0-aproximado}
\end{equation}
We thus obtain, for sufficiently small cavities ($\delta \ll 1$),
\begin{eqnarray}
\left| f_{00}(t)\right| ^2 &\approx &\left( 1+\frac 23\pi \delta
\right) ^{-2}\left\{ 1+\frac{8\delta }\pi \sum_{k=1}^\infty \frac
1{k^2}\cos \left[ \bar{\omega}\left( 1-\frac{\pi \delta }3\right)
-\frac g\delta \left( k+
\frac{2\delta }{\pi k}\right) \right] t\right.   \nonumber \\
&& +\left. \frac{16\delta ^2}{\pi ^2}\sum_{k,l=1}^\infty \frac
1{k^2l^2}\cos \left[ \left( \frac g\delta -\frac{2g}{\pi kl}\right)
(k-l)\right] t\right\} .  \label{rho11}
\end{eqnarray}
To order $\delta ^2$, a lower bound for $\left| f_{00}(t)\right| ^2$ is
obtained by taking the value $-1$ for both cosines in the above formula,
using the tabulated value of the Riemann zeta function $\zeta (2)=\pi ^2/6$,
\begin{equation}
\left| f_{00}(t)\right| ^2\gtrsim \left( 1+\frac 23\pi \delta \right)
^{-2}\left\{ 1-\frac{4\pi \delta }3-\frac{4\pi ^2\delta ^2}9\right\} .
\label{rho11ss}
\end{equation}

We see that the quantity $\left| f_{00}(t)\right| ^2$, which
dictates the behavior of the density matrix elements and of the
measure of purity in equation~(\ref{Dezao1}), has very different
behaviors for free space or for a small cavity. This implies that in
the situation of a small cavity, in contrast to the free space case,
all matrix elements in equations~(\ref{elementos}) are different
from $zero$ for all times.

\begin{figure}[th]
\begin{center}
\includegraphics[{height=4.0cm,width=6.0cm}]{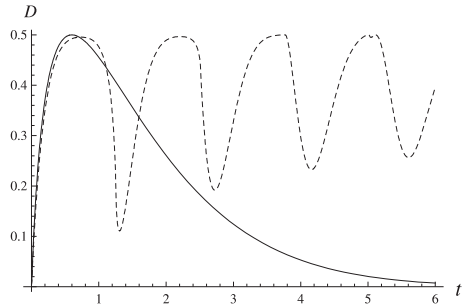}
\end{center}
\caption{Behavior of the degree of impurity $D$ as function of time,
equation (\ref{Dezao1}), for a small cavity (dashed line) and a very large
cavity (solid line); we take the parameters $g=0.5$, $\delta =0.1$
and $\bar{\omega} = 1.0$ (in arbitrary units).} \label{figDezao1}
\end{figure}

In Figure~(\ref{figDezao1}) the degree of impurity from
equation~(\ref{Dezao1}) is plotted as a function of time in the
cases of an arbitrarily large cavity ($R\rightarrow \infty $) and of
a small cavity. We take $\delta =0.1$, with $\bar{\omega}=1.0$ and
$g=0.5$ fixed (in arbitrary units).

We see from the figure that for a very large cavity (free space) the two-atom system dissipates; with the passing of time, both atoms go to their
ground states. For a small cavity the system never completely decay. 

\section{Time evolution of the entanglement}

In order to study how the entanglement of the two-atom states
evolves in time, we shall, as before, consider the system as a
bipartite system, in which each atom carries its own dressing field.
In this way, we shall determine the time behavior of the von Neumann
entropy associated with the reduced density matrix with respect to
one of the subsystems, which is given by taking the trace over the
states of the complementary subsystem.

Let us initially treat the system at $t=0$. Then we have, for the
subsystem $ \mathcal{A}$, the reduced density matrix
\begin{eqnarray}
\rho _{\mathcal{A}}(0) &=&\mathrm{Tr}{}_{\mathcal{B}}\left( \left| \Psi
\right\rangle \left\langle \Psi \right| \right)  \nonumber \\
&=&\sum_{n_{\mathcal{B}},q_j=0}^\infty \left\langle n_{\mathcal{B}%
},q_1,q_2,\ldots \left| \Psi \right. \right\rangle \left\langle \left. \Psi
\right| n_{\mathcal{B}},q_1,q_2,\ldots \right\rangle  \nonumber \\
&=&\xi \left| 1_A,0,0,\cdots \right\rangle \left\langle 1_A,0,0,\cdots
\right| +\left( 1-\xi \right) \left| 0_A,0,0,\cdots \right\rangle
\left\langle 0_A,0,0,\cdots \right|  \label{matriz densidade A}
\end{eqnarray}
and, similarly, for the subsystem $\mathcal{B}$,
\begin{eqnarray}
\rho _{\mathcal{B}}(0) &=&\mathrm{Tr}_{\mathcal{A}}\left( \left| \Psi
\right\rangle \left\langle \Psi \right| \right)  \nonumber \\
&=&\xi \left| 0_B,0,0,\cdots \right\rangle \left\langle 0_B,0,0,\cdots
\right| +\left( 1-\xi \right) \left| 1_B,0,0,\cdots \right\rangle
\left\langle 1_B,0,0,\cdots \right| .
\end{eqnarray}

The degree of entanglement of the two-atom system is measured by the von
Neumann entropy of any of the reduced density matrices; for instance,
\begin{equation}
E(\xi )=-\mathrm{Tr}\left[ \rho _{\mathcal{A}}\ln \rho _{\mathcal{A}}\right]
=-\sum_\alpha \alpha \ln \alpha ,
\end{equation}
where the sum is taken over the eigenvalues $\alpha $ of $\rho
_{\mathcal{A}} $. Since $\rho _{\mathcal{A}}$ is diagonal in the
Fock basis of the dressed states of the atom $\mathcal{A}$, its
eigenvalues can be read directly from (\ref{matriz densidade A}):
\begin{equation}
\alpha _1 = 1-\xi ,\qquad \alpha _2=\xi ,\qquad \alpha _2=\alpha
_3=\cdots =0.
\end{equation}
Therefore,
\begin{equation}
E(\xi )=-\left[ (1-\xi )\ln (1-\xi )+\xi \ln \xi \right] .  \label{entropy}
\end{equation}

The time evolution of the states $\left| \Gamma _{1(A)0(B)00\cdots
}^{(AB)}\right\rangle $ and $\left| \Gamma _{0(A)1(B)00\cdots
}^{(AB)}\right\rangle $ are governed by the time evolution of the
states $ \left| \Gamma _{100\cdots }^A\right\rangle $ and $\left|
\Gamma _{100\cdots }^B\right\rangle $, respectively, given by
equation~(\ref{ortovestidas51}),
\begin{equation}
\left| \Gamma _{100\cdots }^\lambda (t)\right\rangle =\sum_\nu f_{\lambda
\nu }(t)\left| \Gamma _{100\cdots }^{\nu (\lambda )}\right\rangle ,
\label{ortovestidas52}
\end{equation}
where, in accord with the notation of the preceding section, the
label $ \lambda $ now refers to each one of the dressed atoms $A$
and $B$ and
\begin{equation}
f_{\lambda \nu }(t)=\sum_st_\lambda ^st_\nu ^se^{-i\Omega _st}.
\label{ortovestidas53}
\end{equation}
In equation~(\ref{ortovestidas52}), $\left| \Gamma _{100\cdots }^{\nu
(\lambda )}\right\rangle $ is the state in which the dressed mode $\nu
(\lambda )$ of the atom $\lambda $ is at the first level and all the other
dressed modes are in the ground state.

The reduced density matrix corresponding to the subsystem $\mathcal{A}$ at
time $t$ is
\[
\rho _{\mathcal{A}}(t)=\mathrm{Tr}_{\mathcal{B}}\varrho
(t)=\mathrm{Tr}_{\mathcal{B}}\left[ \left| \Psi (t)\right\rangle
\left\langle \Psi (t)\right| \right] .
\]
Using equation (\ref{ortovestidas52}), one writes $\rho _{\mathcal{A}}(t)$
in terms of the quantities $f_{\lambda \nu }(t)$ from equation (\ref
{ortovestidas53}):
\begin{eqnarray}
\rho _{\mathcal{A}}(t) &=&\sum_{n_{\mathcal{B}},q_j=0}^\infty \left\langle
n_{\mathcal{B}},q_1,q_2,\ldots \left| \Psi (t)\right. \right\rangle
\left\langle \left. \Psi (t)\right| n_{\mathcal{B}},q_1,q_2,\ldots
\right\rangle  \nonumber \\
&=&\sum_{n_{\mathcal{B}},q_j}\sum_{\mu ,\nu }\left[ \sqrt{\xi
}f_{A\mu }(t)\left| \Gamma _{100\cdots }^{\mu (A)}\right\rangle
\left\langle n_{ \mathcal{B}},q_1,q_2,\ldots \left| \Gamma
_{000\cdots }^B\right. \right\rangle +\sqrt{1-\xi }\,e^{i\phi
}f_{B\mu }(t)\left| \Gamma _{000\cdots }^A\right\rangle \left\langle
n_{\mathcal{B}},q_1,q_2,\ldots \left| \Gamma _{100\cdots }^{\mu
(B)}\right. \right\rangle \right]
\nonumber \\
&&\times \left[ \sqrt{\xi }f_{A\nu }^{*}(t)\left\langle \Gamma
_{100\cdots }^{\nu (A)}\right| \left\langle \left. \Gamma
_{000\cdots }^B\right| n_{ \mathcal{B}},q_1,q_2,\ldots \right\rangle
+\sqrt{1-\xi }\,e^{-i\phi }f_{B\nu }^{*}(t)\left\langle \Gamma
_{000\cdots }^A\right| \left\langle \left. \Gamma _{100\cdots }^{\nu
(B)}\right| n_{\mathcal{B}},q_1,q_2,\ldots
\right\rangle \right]   \nonumber \\
&=&\sum_{\mu ,\nu }\xi f_{A\mu }(t)f_{A\nu }^{*}(t)\left| \Gamma _{100\cdots
}^{\mu (A)}\right\rangle \left\langle \Gamma _{100\cdots }^{\nu (A)}\right|
+(1-\xi )\left| \Gamma _{000\cdots }^A\right\rangle \left\langle \Gamma
_{000\cdots }^A\right| ,  \label{roa}
\end{eqnarray}
where we have used
\begin{equation}
\sum_{\mu(\lambda)} f_{B\mu }(t)\left\langle n_{\mathcal{B}},q_1,q_2,\ldots
\left| \Gamma _{100\cdots }^{\mu (B)}\right. \right\rangle
=f_{BB}(t)\delta _{n_{ \mathcal{B}}1}\prod_i\delta _{i0}+\delta
_{n_{\mathcal{B}}0}\sum_if_{Bi}(t) \delta _{i1}\prod_{j\neq i}\delta
_{j0}
\end{equation}
and equation (\ref{probabilidade}).

\begin{figure}[th]
\begin{center}
\includegraphics[{height=4.0cm,width=6.0cm}]{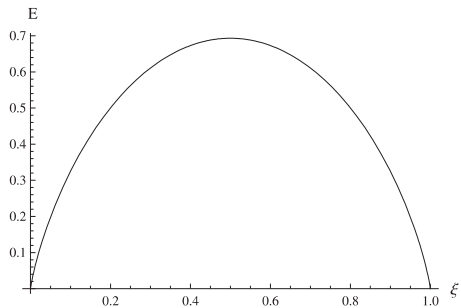}
\end{center}
\caption{Behaviour of the von Neumann entropy at all times, equation (\ref
{entropy-t}), as a function of the parameter $\xi $.}
\label{G}
\end{figure}

The time-dependent von Neumann entropy is now given by
\begin{equation}
E(t,\xi )=-\mathrm{Tr}\left[ \rho _{\mathcal{A}}(t)\ln \rho
_{\mathcal{A} }(t)\right] =-\sum_\alpha \alpha \ln \alpha ,
\end{equation}
where here $\alpha $ are the time-dependent eigenvalues of the reduced
density matrix. These should be solutions of the so-called characteristic
equation, which in the case of (\ref{roa}), reads
\begin{equation}
\det \left(
\begin{array}{ccccc}
1-\xi -\alpha  &  &  &  &  \\
& \xi \left| f_{AA}\right| ^2-\alpha  & \xi f_{A1}f_{AA}^{*} & \xi
f_{A2}f_{AA}^{*} & \cdots  \\
& \xi f_{AA}f_{A1}^{*} & \xi \left| f_{A1}\right| ^2-\alpha  & \xi
f_{A2}f_{A1}^{*} & \cdots  \\
& \xi f_{AA}f_{A2}^{*} & \xi f_{A1}f_{A2}^{*} & \xi \left| f_{A2}\right|
^2-\alpha  & \cdots  \\
& \vdots  & \vdots  & \vdots  & \ddots
\end{array}
\right) =0.
\end{equation}
We thus find that the nonzero eigenvalues of $\rho _{\mathcal{A}}$ are
\begin{equation}
\alpha _1=1-\xi ,\qquad \alpha _2=\xi \sum_{\mu(\lambda)} \left| f_{A\mu }(t)\right|
^2=\xi .
\end{equation}
This then implies that the von Neumann entropy takes the expression$\ $
\begin{equation}
E(t,\xi )=-\left[ (1-\xi )\ln (1-\xi )+\xi \ln \left( \xi \right) \right] ,
\label{entropy-t}
\end{equation}
that is, all the time dependence of the von Neumann entropy for this
two-atom system, coming from the $f_{\lambda \nu }(t)$, is completely
cancelled in the computation of the entropy, in all situations, thereby
reproducing exactly the same expression as in the $t=0$ case, with the
maximum entanglement occuring at $\xi =1/2$ (see Figure \ref{G}). In other
words, although the superposition of states evolves in time, in different
ways in the limits of a very large cavity and of a small one, the entangled
nature of these two-atom states remains unchanged for all times,
independently of the size of the cavity.

\section{Concluding remarks}

In this paper we have considered a system composed of two atoms in a
spherical cavity, each of them in independent interaction with an
environment field. The model employed is of a bipartite system, in which
each subsystem consists of one of the atoms dressed by the field. We make
the assumption that initially we have a state in which one of the dressed
atoms is in its first excited level and the other atom and the field modes
are all in the ground state, is superposed with a state in which the atoms
have their roles reversed.

The time evolution of the superposed states leads to a
time-dependent (reduced) density matrix. Expressions for its
elements are provided in both the cases of an infinitely large
cavity (that is, free space) and of a small one, when the two atoms
are considered as identical. Very different behaviors are obtained
for this time evolution. In the large-cavity case, the system shows
dissipation, and, with the passing of time, both atoms go to their
ground states. For a small cavity, an oscillating behavior is
present, so that the atoms never fully decay.

Nevertheless, in spite of these rather contrasting behaviors and of the
nontrivial time dependence of the density matrix, we obtain a
time-independent von Neumann entropy, which means that the initial
entanglement of the two atoms remains unchanged as the system evolves.
\\

\noindent \textbf{Acknowledgments:} The authors acknowledge CAPES and
CNPq/MCT (Brazil) for partial financial support.

\end{document}